\documentclass[%
nofootinbib,
 amsmath,amssymb,
 aps,
twocolumn,
]{revtex4-1}

\usepackage[utf8]{inputenc}
\usepackage[english]{babel}
\usepackage{titlesec}
\usepackage{etoolbox}
\usepackage{amsmath}
\usepackage{amssymb} 

\usepackage{graphicx} 
\usepackage{xcolor}
\usepackage{natbib}
\usepackage{braket}

\usepackage{siunitx}

\usepackage{cancel}
\newcommand\Ccancel[2][black]{
    \let\OldcancelColor\CancelColor
    \renewcommand\CancelColor{\color{#1}}
    \cancel{#2}
    \renewcommand\CancelColor{\OldcancelColor}
}

\newcommand{\inv}{^{-1}}

\usepackage[normalem]{ulem}
\newcommand{\stkout}[1]{\ifmmode\text{\sout{\ensuremath{#1}}}\else\sout{#1}\fi}

\usepackage[all]{nowidow}

\begin{document}

\preprint{APS/123-QED}

\title{Enhanced Sampling of Configuration and Path Space in a Generalized Ensemble by Shooting Point Exchange}
\date{\today}

\author{Sebastian Falkner 
}
\affiliation{University of Vienna, Faculty of Physics, 1090 Vienna, Austria.}

\author{Alessandro Coretti 
}
\affiliation{University of Vienna, Faculty of Physics, 1090 Vienna, Austria.}

\author{Christoph Dellago 
}
\email{christoph.dellago@univie.ac.at}
\affiliation{University of Vienna, Faculty of Physics, 1090 Vienna, Austria.}

\begin{abstract}

The computer simulation of many molecular processes is complicated by long time scales caused by rare transitions between long-lived states. Here, we propose a new approach to simulate such rare events, which combines transition path sampling with enhanced exploration of configuration space. The method relies on exchange moves between configuration and trajectory space, carried out based on a generalized ensemble. This scheme substantially enhances the efficiency of the transition path sampling simulations, particularly for systems with multiple transition channels, and yields information on thermodynamics, kinetics and reaction coordinates of molecular processes without distorting their dynamics. The method is illustrated using the isomerization of proline in the KPTP tetrapeptide.

\end{abstract}

\maketitle

Overcoming high energy barriers to explore configuration and trajectory space in simulations of rare events is at the core of the sampling problem. Numerous enhanced sampling techniques have been developed over the years to better understand the thermodynamics and kinetics of rare events such as nucleation, chemical reactions and biomolecular reorganization~\cite{Arjun2019,Menzl2016a,Leitold2020,Juraszek2006,Okazaki2019}. However, when deciding for the most suitable method, a conflict of interest frequently arises. Enhanced sampling methods such as metadynamics~\cite{Laio2002,Barducci2008} and umbrella sampling~\cite{TorrieG.MValleau1977} efficiently focus computational resources on the regions of interest while still allowing for reweighting to gain information on the equilibrium distribution $p_{\text{eq}}(x)$. However, the dynamics of the system are distorted by the introduction of the bias potential. In contrast, path sampling methods such as Transition Path Sampling (TPS) and Transition Interface Sampling (TIS) allow to obtain true-dynamic trajectories between stable states~\cite{Csajka1998,Bolhuis2002}. However, these schemes may suffer from correlations between subsequently sampled trajectories, in particular for systems with several reaction channels.
To alleviate this problem, in recent years there has been great interest in applying enhanced sampling methods to path space~\cite{Bolhuis2018,Borrero2016,Mandelli2020,Falkner2022}. Apart from that, points on the trajectory are not distributed according to $p_{\text{eq}}(x)$. Although in principle possible, it is unfeasible to reweight samples to the equilibrium distribution since this requires knowledge of the committor probability of each point~\cite{Hummer2004}.

In this letter, we propose a sampling scheme based on a parallel sampling of configuration and path space, e.g., using metadynamics and TPS. The two simulations are coupled by exchanging configurations between them following an acceptance criterion derived for a generalized ensemble. As a result, transition paths show less correlations due to fast relaxation in configuration space and barriers in configuration space are crossed more frequently due to exchanges with configurations on transition paths.

Exchange moves, e.g. as employed in replica exchange molecular dynamics~\cite{Swendsen1986,Sugita1999} or replica exchange TIS~\cite{vanErp2007}, are a powerful tool to enhance sampling and reduce correlations between samples. In such a move~\cite{GilLey2015}, configurations of two systems are exchanged according to an acceptance criterion to ensure a properly weighted ensemble in both systems in the limit of infinite sampling. We propose to perform exchange moves between configurations $y$ from a given distribution $p_y(y)$ and a configuration on a path $X$ from the distribution of transition paths $P_X^\text{AB}(X)$, which includes only paths that connect two given regions A and B. In this context, a generalized ensemble can be defined through the joint distribution:
\begin{align}
\label{eq:generalizedEnsemble}
P_z(z) =  p_y(y)\,P^{\text{AB}}_X \left[ X(\tau) \right]
\end{align}
where the state $z$ is given by $z = \{y, X(\tau)\}$ with a configuration $y$ and a path $X(\tau)$ of length $\tau$. The probability density of reactive paths $P^{\text{AB}}_X \left[ X(\tau) \right]$ is given by~\cite{Bolhuis2002}:
\begin{align}
    \label{reactive_path_probability}
    P^{\text{AB}}_X \left[ X(\tau) \right] =&
    \frac{1}{Z_{\text{AB}}}\,
    H_{\text{AB}}(x_0, x_\tau)  
    \prod_{i=1}^{\tau/\Delta t-1}  \widetilde{h}(x_{i \Delta t})\  \notag \\
    & \times P_X \left[ X(\tau) \right]
\end{align}
where $\Delta t$ is the timestep, $p(x_{i\Delta t} \to x_{(i+1)\Delta t})$ is the short-time transition probability from $x_{i\Delta t}$ to $x_{(i+1)\Delta t}$ and $Z_\text{AB}$ is the partition function. The probability distribution of an unconstrained path $P_X \left[ X(\tau) \right]$ is given by~\cite{Bolhuis2002}:
\begin{align}
    \label{dynamical_path_probability}
    P_X \left[ X(\tau) \right] =& p_\text{eq}(x_0) \prod_{i=0}^{\tau/\Delta t-1} p(x_{i\Delta t} \to x_{(i+1)\Delta t})\ .
\end{align}
where $p_\text{eq}(x)$ is the equilibrium or stationary distribution for the underlying dynamics of $X$. The factor $H_{\text{AB}}(x_0, x_\tau)$ is one if the trajectory connects states A and B in any order and is zero otherwise. The function $\widetilde{h}(x)$ is zero if $x$ is in state A or B and unity otherwise, ensuring that the transition path has exactly one point in state A and one in B.

\begin{figure}
    \centering
    \includegraphics[width=\columnwidth]{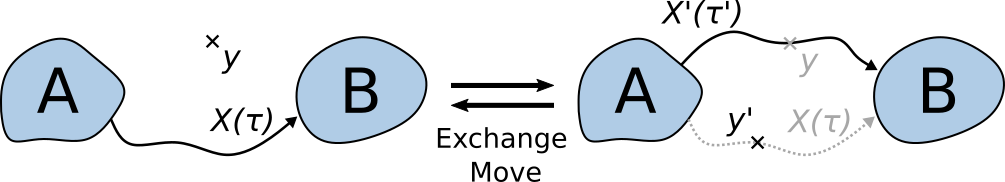}
    \caption{Schematic representation of the proposed exchange move for the generation of a new path $X'(\tau')$ and configuration $y'$ given the path $X(\tau)$ and configuration $y$.}
    \label{fig:exchange_scheme}
\end{figure}

For an exchange between the two spaces in $P_z(z)$, Eq.~\eqref{eq:generalizedEnsemble}, a new state $z' = \{y', X'(\tau')\}$ is generated based on the current state $z = \{y, X(\tau)\}$ (Figure~\ref{fig:exchange_scheme}). The new configuration $y'$ is obtained by selecting a point on the current trajectory $X(\tau)$ with probability $p^\text{sel} \bigl[ y' | X(\tau)\bigr]$. The generation probability for this move is given by:
\begin{align}
\label{Eq:genPrb_y}
p^\text{gen}_y \bigl[ X(\tau) \to y'\bigr] = p^\text{sel} \bigl[ y' | X(\tau)\bigr]
\end{align}
A new path $X'$ is generated based on $y$ by means of a shooting move where the equations of motion are integrated forward and backward in time until a stable state is reached. With $k \Delta t$ being the time of the shooting point on the new path, the generation probability is given by:
\begin{align}
P^\text{gen}_X \bigl[ y \to X'(\tau') \bigr]  =&  
    \prod_{i=k}^{\tau'/\Delta t-1} p(x'_{i\Delta t} \to x'_{(i+1)\Delta t}) \notag\\
   &\times\prod_{i=1}^{k} \bar{p}(x'_{i\Delta t} \to x'_{(i-1)\Delta t}) .
\end{align}
Assuming that the transition probabilities fulfill microscopic reversibility, the above distribution can be rewritten as:
\begin{align}
\label{Eq:genPrb_X}
P^\text{gen}_X \bigl[ y \to X'(\tau') \bigr] = \frac{1}{p_\text{eq}(y)} 
\times P_X \left[ X'(\tau') \right]
\end{align}
Imposing detailed balance, the acceptance probability for the described exchange move must obey
\begin{align}
\frac{P^\text{acc}_z \bigl( z \to z' \bigr)}{P^\text{acc}_z \bigl( z' \to z \bigr)} = \frac{P_z \bigl( z'\bigr) P^\text{gen}_z \bigl( z' \to z \bigr)}{P_z \bigl( z\bigr) P^\text{gen}_z \bigl( z \to z' \bigr)}
\end{align}
which can be satisfied using the Metropolis rule;
\begin{align}
P^\text{acc}_z \bigl( z \to z' \bigr) = \min \Biggl\{ 1, \frac{P_z \bigl( z'\bigr) P^\text{gen}_z \bigl( z' \to z \bigr)}{P_z \bigl( z\bigr) P^\text{gen}_z \bigl( z \to z' \bigr)}\Biggr\} 
\end{align}
Inserting expressions from Eq.~\eqref{eq:generalizedEnsemble}, \eqref{Eq:genPrb_y} and \eqref{Eq:genPrb_X}, the acceptance criterion for the exchange move is:
\begin{align}
\label{eq:acceptance}
P^\text{acc}_z \bigl( z \to& z'  \bigr)
= H_{\text{AB}}(x'_0, x'_{\tau'}) \prod_{i=1}^{\tau'/\Delta t-1} \widetilde{h}(x'_{i \Delta t}) \notag\\
&\times \min \Biggl\{ 1, 
\frac{p_y(y')}{p_y(y)}
\frac{p_\text{eq}(y)}{p_\text{eq}(y')}
\frac{p^\text{sel} \bigl[ y | X'(\tau')\bigr]}{p^\text{sel} \bigl[ y' | X(\tau)\bigr]}
\Biggr\}
\end{align}

The exchange scheme is most efficient if $p_y(y)$ has significant overlap with $p_\text{eq}(y)$ and shooting moves starting from samples of $y$ have a reasonable probability of generating a transition path. Both of these conditions can be matched well by setting the distribution $p_y(y)$ to the Boltzmann distribution with a bias potential introduced via metadynamics~\cite{Laio2002,Barducci2008}:
\begin{align}
p_y(y) = Z\inv \exp \bigl\{-\beta [U(y) + U_{\text{bias}}(r(y))]\bigr\}
\end{align}
The acceptance of the exchange scheme can then be further improved by tuning the selection probability $p^\text{sel} \bigl[ y' | X(\tau)\bigr]$, which represents the probability to choose a point $y'$ on a given path $X(\tau)$. We can bias this selection in the spirit of Jung et al.~\cite{Jung2017} according to the current bias introduced by metadynamics:
\begin{align}
p^\text{sel}\bigl[ y' | X(\tau)\bigr] = \frac{\exp \bigl\{-\beta [U_{\text{bias}}(r(y'))]\bigr\}}
{ \sum_{i=0}^{\tau/\Delta t} \exp \bigl\{-\beta [U_{\text{bias}}(r(x_{i \Delta t}))]\bigr\}}
\end{align}
The acceptance probability for the exchange then becomes:
\begin{align}
\label{eq:acceptance_simple}
P^\text{acc}_z&\bigl( z \to z' \bigr) =H_{\text{AB}}(x'_0, x'_{\tau'}) \prod_{i=1}^{\tau'/\Delta t-1} \widetilde{h}(x'_{i \Delta t}) \notag\\
&\times \min \Biggl\{ 1, 
\frac{
\sum_{i=0}^{\tau/\Delta t} \exp \bigl\{-\beta [U_{\text{bias}}(r(x_{i \Delta t}))]\bigr\}
}{
\sum_{i=0}^{\tau'/\Delta t} \exp \bigl\{-\beta [U_{\text{bias}}(r(x'_{i \Delta t}))]\bigr\}
}\Biggr\}
\end{align}

The resulting criterion therefore represents the ratio of the times the old and new paths spend in regions with a high bias potential. This expression is very similar to the reweighting factor necessary when initiating paths from a biased distribution of shooting points~\cite{Falkner2022}. During the exchange, $y$ acts as a shooting point to generate $X'$ and $y'$, selected on $X$, is chosen with the same procedure as a shooting point in regular TPS. Therefore, we call the exchange scheme \emph{shooting point exchange} (SPEx) in the following. 

\begin{figure*}
    \centering
    \includegraphics[width=\textwidth]{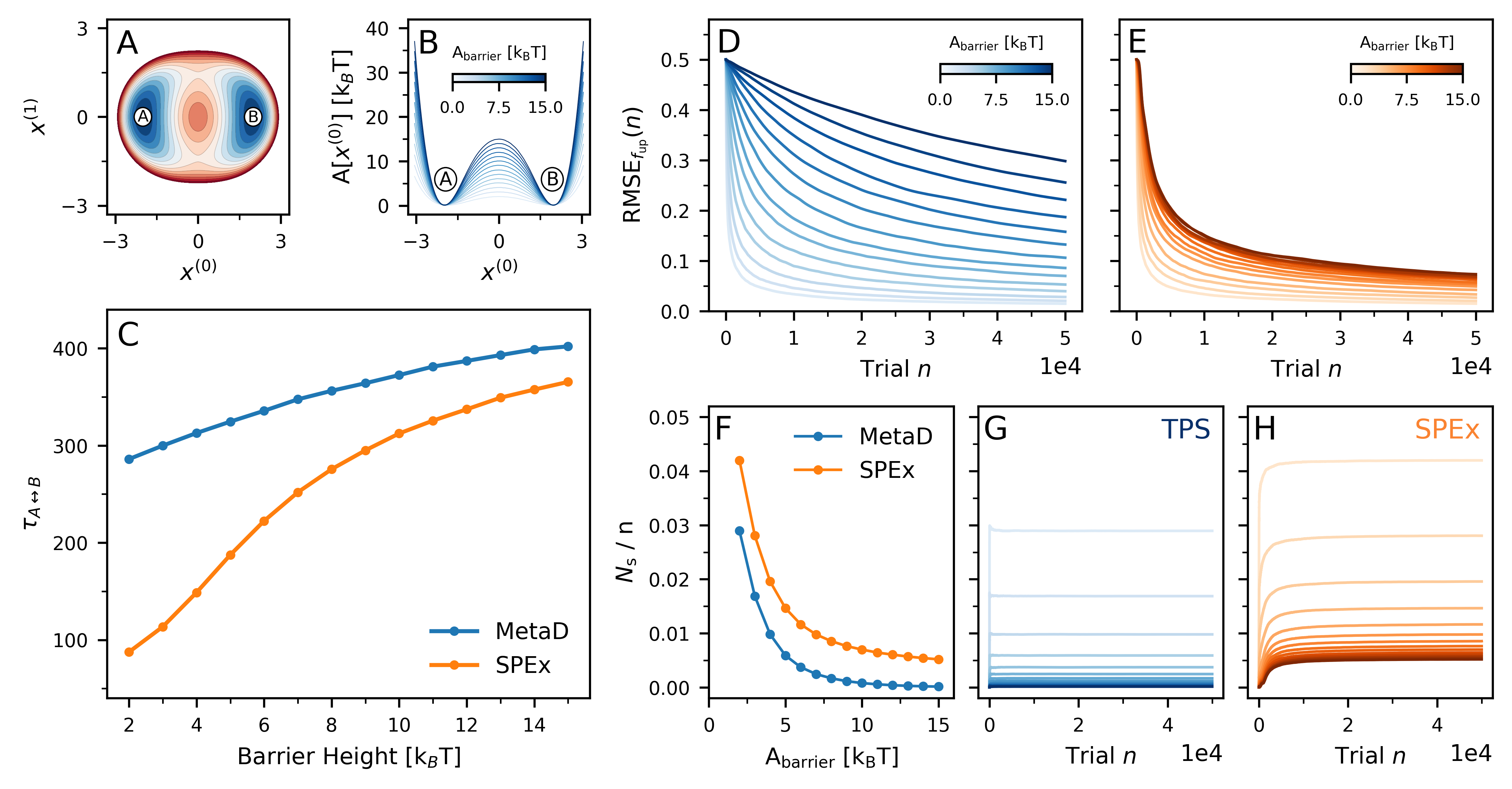}
    \caption{Shooting point exchange for sampling the two reaction channels in the two-dimensional double well model with two reaction channels. (A) Potential energy and state definitions of the model system. (B) Free energy as a function of $x^{(0)}$ for different barrier heights. (C) Average switching time to observe a switch between states A and B, $\tau_{\text{A} \leftrightarrow \text{B}}$, for standalone metadynamics and SPEx, starting the sampling from a converged bias potential. (D, E) Root mean squared error of the fraction of paths in the upper reaction channel as a function of simulation length for standalone TPS and SPEx. Each curve is for a specific barrier height and is estimated from $2500$ independent sampling runs. (F,G,H) Number of switches between the upper and lower reaction channel $N_s$ as a function of the barrier height (F) and the number of trials (G,H).}
    \label{fig:BSDW}
\end{figure*}

\begin{figure*}
    \centering
    \includegraphics[width=\textwidth]{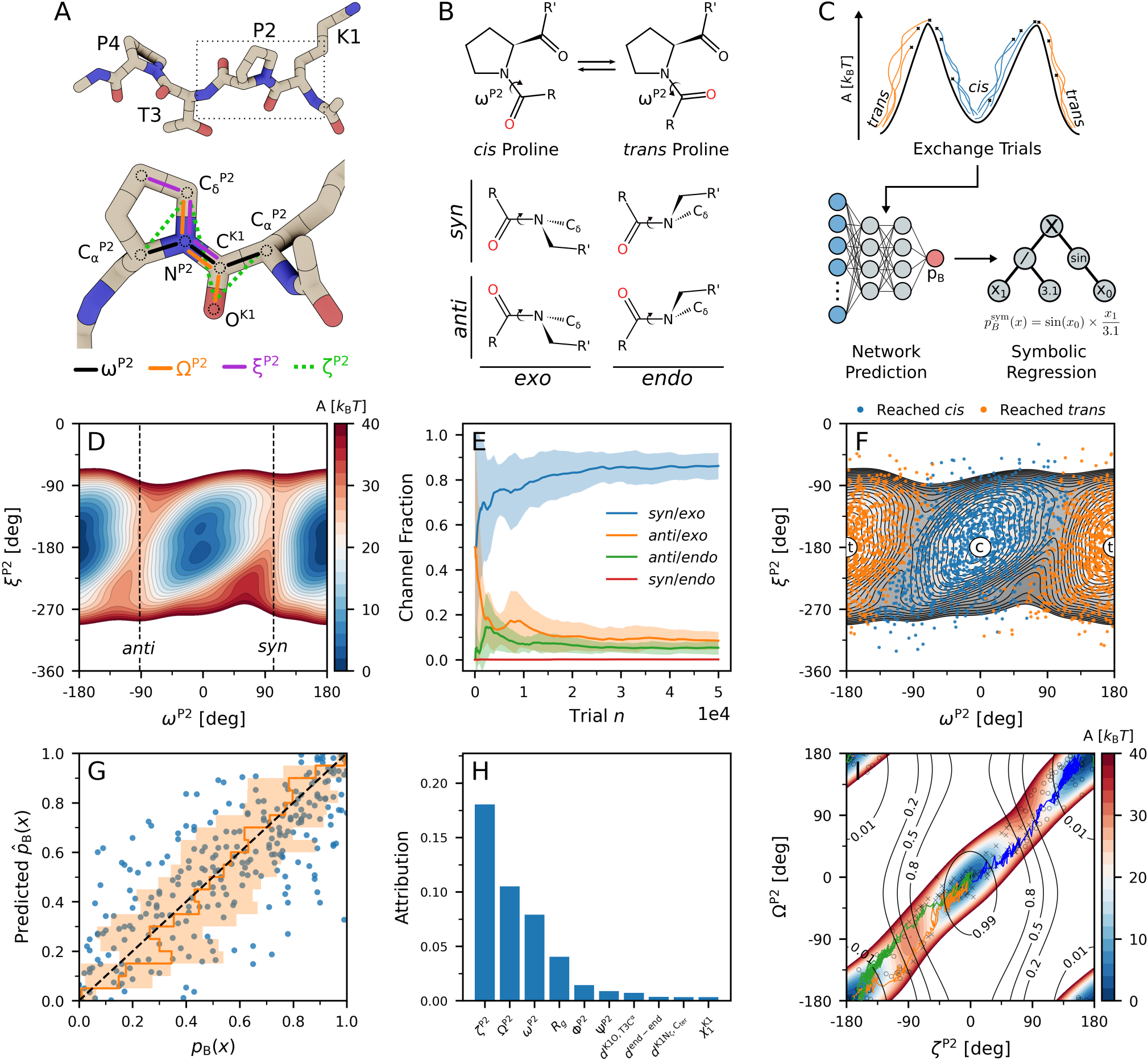}
    \caption{Isomerization of proline in the tetrapeptide KPTP. (A) Structure of the peptide (relevant torsion angle definitions in the inset) (B) Scheme of the \textit{cis} to \textit{trans} isomerization and the transition state geometry (C) Schematic overview of the committor learning process. (D) Free energy from metadynamics including shooting point exchanges. (E) Fraction of paths in each reaction channel as a function of the number of trials based on ten independent shooting point exchange simulations. (F) Training data for the committor prediction on top of the free energy surface. Circles show the state definitions (c=\textit{cis}, t=\textit{trans}). (G) Comparison of the sampled committor from fleeting trajectories and the predicted committor. The dashed black line shows the ideal correspondence while the orange line and shaded area show the average and standard deviation of the sampled committor in a given window of the predicted committor. (H) Attributions corresponding to the ten most important input features of the neural network. (I) Free energy along the two most relevant collective variables from (H). Black lines show isolines of the committor function obtained using symbolic regression. Crosses and circles indicate if a trajectory starting from that point reached the \textit{cis} or \textit{trans} state first. Representative reactive paths are shown in the same color scheme as in (E).}
    \label{fig:KPTP}
\end{figure*}

We first test the sampling scheme on a double well model, where stable states are connected by two distinct reaction channels (Figure~\ref{fig:BSDW}A,B). In this system, two factors complicate the efficient sampling of configuration and path space. On the one hand, stable states are separated by an energy barrier limiting the occurrence of switches between them. On the other hand, also the two reaction channels are separated by a barrier. Hence, sampling transition paths connecting states A and B suffers from strong correlations since subsequently visited paths tend to remain in the same reaction channel. We compare the performance of standalone metadynamics and TPS with the performance of SPEx for different heights of the barrier separating the stable states (simulation details in supplementary information, SI~\cite{SI}). In configuration space, sampled with metadynamics, we measure the time needed to switch between stable states A and B, $\tau_{\text{A} \leftrightarrow \text{B}}$. The exchange moves decrease the switching time between the two states at all barrier heights (Figure~\ref{fig:BSDW}C). As the speedup is linked to the number of accepted exchanges, the effect is more pronounced at smaller barrier heights due to an increased likelihood to generate a transition path also from configurations away from the barrier. 

For transition paths, a limiting factor for the sampling of trajectories that connect A and B is the slow switching between the upper and lower reaction channel. In regular TPS, the fraction of paths taking the upper reaction channel converges very slowly to the analytical fraction of $f_\text{up}=\frac{1}{2}$. This is apparent looking at the root mean square error of $f_\text{up}$ as a function of the trial number $n$ estimated from $N$ runs (Figure~\ref{fig:BSDW}D,E): 
\begin{align}
\text{RMSE}_{f_\text{up}}(n) = \sqrt{\sum_{i=0}^N \biggl(f_\text{up}(n) - \frac{1}{2}\biggr)^2}
\end{align}

In comparison, propagation of the system using metadynamics with shooting point exchange speeds up the convergence substantially (Figure~\ref{fig:BSDW}E), especially for high barriers. This can be traced back to an increased number of switches during the sampling of the path ensemble (Figure~\ref{fig:BSDW}F,G,H). Above barriers of $10\,k_\text{B}T$, often not even a single switch between the channels occurs within $1000$ trials in standalone TPS. When exchange moves are included, a minimum number of switches is recovered, mostly mediated by the exchange moves themselves.

From these observations, we can conclude that exchanges between an enhanced sampling simulation in configuration space and a path sampling simulation increase the sampling efficiency on either side compared to running both simulations separately. While the increased switching time $\tau_{\text{A} \leftrightarrow \text{B}}$ is not negligible, the additional cost of sampling a path ensemble in parallel is not compensated. Therefore, at least in this setup, SPEx is not expected to increase the sampling efficiency when the interest is solely on configuration space. In contrast, when sampling path space, the additional force evaluations from metadynamics per TPS trial are insignificant compared to the number of force evaluations needed for the generation of a new path.

As a second case study, we investigate the \textit{cis} to \textit{trans} isomerization of the amino acid proline in the tetrapeptide KPTP~\cite{Alcantara2021} (Figure~\ref{fig:KPTP}A). Proline isomerization plays an important role in protein folding~\cite{Wedemeyer2002,Favretto2020} and signaling in cells~\cite{Lu2007,Sarkar2007}, yet it only occurs on the timescale of seconds to minutes~\cite{Grathwohl1981}. Due to the periodic nature of the relevant imide torsion angle $\omega^\text{P2}$, the transition from \textit{cis} to \textit{trans} and vice versa can take place via different reaction channels (Figure~\ref{fig:KPTP}B). During the transition from \ang{\pm180} (\textit{trans}) to \ang{0} (\textit{cis}), the torsion angle can either cross over a barrier at \ang{-90} or \ang{108}, referred to as \textit{anti} and \textit{syn} conformation~\cite{Fischer1994} (dashed lines in figure~\ref{fig:KPTP}D). Additionally, the imide nitrogen geometry, which is planar in the stable \textit{cis} and \textit{trans} states, is deformed out of plane~\cite{Fischer1994}. By the direction of the deformation indicated by the torsion angle $\eta_\text{P2}$, the transition state can be distinguished as \textit{endo} or \textit{exo}, resulting in a total of four channels. The critical out of plane deformation is not captured by $\omega^\text{P2}$ and therefore previous works proposed an improper dihedral $\zeta_\text{P2}$ as reaction coordinate~\cite{Fischer1994,Melis2009} (Figure~\ref{fig:KPTP}A). Other collective variables discussed in the context of the isomerization are the $\Psi_\text{P2}$ backbone angle~\cite{Fischer1994,Melis2009,Martino2014}, the puckering state~\cite{Kang2004,Wu2013} of the ring and solvent interactions~\cite{Ke1993}.

Previous studies that focused on the mechanism of proline isomerization mainly used biased molecular dynamics to enhance the sampling~\cite{Alcantara2021,Melis2009,Martino2014,Kang2004,Wu2013}. As a result, the dynamics of the system was altered and conclusions on the preferred mechanism and a corresponding reaction coordinate were mostly drawn based on minimum energy paths, with a notable exception being the recent work by Moritsugu et al. on the Pin1 enzyme~\cite{Moritsugu2021}. We aim to identify the preferred isomerization mechanism, search systematically for relevant degrees of freedom and refine a reaction coordinate based on the unbiased dynamics of the system (simulation details in SI~\cite{SI}). Besides the imide torsion angle, we choose $\xi^\text{P2}$ (see Figure~\ref{fig:KPTP}A) for sampling, as we expected it to capture potential geometric changes of both the imide nitrogen and the sidechain. The resulting free energy from metadynamics with shooting point exchanges agrees with previous studies in terms of the barrier height and difference between the \textit{cis} and \textit{trans} state~\cite{Alcantara2021,Martino2014} (Figure~\ref{fig:KPTP}D). However, estimating the different statistical weights of the four reaction channels is not possible. The \textit{endo} and \textit{exo} paths are not discriminated by $\xi^\text{P2}$ and, more importantly, an estimation based on barrier heights does not account for entropy in path space. Here, the sampled transition paths can give an accurate estimate of the fraction of paths going through each channel (Figure~\ref{fig:KPTP}E). These are \textit{syn/exo} $0.862$, \textit{anti/exo} $0.084$, \textit{anti/endo} $0.053$ and \textit{syn/endo} $0.001$, pointing out a clear preference for the \textit{syn/exo} pathway.

Intending to find an improved reaction coordinate for the transition, we train a neural network to predict the committor probability $p_\text{B}(x)$ of a given configuration as proposed by Jung et al.~\cite{Jung2021}. The committor describes the likelihood of reaching state B before state A starting a simulation from configuration $x$ and thereby also describes the progress of a reaction. From a broad set of collective variables based on which the network predicts the committor, the most important features can be determined by assigning an attribution score~\cite{Jung2021}. In the context of SPEx, we train the network using the information obtained from exchange moves (Figure~\ref{fig:KPTP}F, network details in SI~\cite{SI}). Although the training data only contain labels indicating if \textit{cis} or \textit{trans} was reached first, the network learns to interpolate in ambiguous regions (Figure~\ref{fig:KPTP}G). Looking at the assigned attributions (Figure~\ref{fig:KPTP}H), the torsion angles $\zeta^\text{P2}$, $\Omega^\text{P2}$ and $\omega^\text{P2}$ are the most important variables, followed by the radius of gyration $R_g$, which has previously been discussed to be linked to the fraction of \textit{cis} proline residues~\cite{Alcantara2021}. Collective variables describing the puckering state of the ring, $\Psi_\text{P2}$ and all other backbone angles do not contribute significantly to the prediction of the committor. The neural network prediction is then used to refine an expression for a reaction coordinate via symbolic regression. Here, we include only the three torsion angles in the analysis in an attempt to obtain a reaction coordinate independent of the peptide sequence. 

The most accurate estimate of $p_\text{B}(x)$ from symbolic regression includes $\zeta^\text{P2}$ and $\Omega^\text{P2}$ (Figure~\ref{fig:KPTP}I):
\begin{align}
    p_\text{B} (\zeta^\text{P2}, \Omega^\text{P2}) = \text{sig}\bigl[&- \sin{\left(\Omega^\text{P2} - 0.75 \right)} + 4.334 \cos{\left(\zeta^\text{P2} \right)} \notag\\
    &+ \cos{\left(\Omega^\text{P2} \right)} - 0.635 \bigr],
\end{align}
where $\text{sig}(x) = 1/[1+\exp(-x)]$. Although $\zeta^\text{P2}$ is undoubtedly a better reaction coordinate than $\omega^\text{P2}$ as shown in previous works, the committor isolines indicate that at least $\Omega^\text{P2}$ is required for an accurate prediction of $p_\text{B}(x)$.

To conclude, we presented a framework based on exchange moves between a configuration and a path ensemble. The sampling scheme has the potential to efficiently explore free energy surfaces, transition path ensembles and reaction coordinates of molecular processes, as demonstrated on the proline \textit{cis}-\textit{trans} isomerization. The case studies presented here --- combining metadynamics and TPS --- are just one realization of the possibilities emerging from Eq.~\eqref{eq:acceptance}. Since the generalized ensemble is not limited to a single configuration and path ensemble, we see future applications e.g. in umbrella sampling~\cite{TorrieG.MValleau1977}, multi-state TPS~\cite{Rogal2008} or TIS~\cite{vanErp2003}.

\section*{Data Availability}
The data that support the findings of this study are available upon reasonable request. 

\begin{acknowledgments}
We acknowledge financial support of the Austrian Science Fund (FWF) through the SFB TACO, Grant number F 81-N. The computational results presented were achieved using the Vienna Scientific Cluster (VSC).
\end{acknowledgments}


%

\clearpage

\onecolumngrid
\normalsize
\patchcmd{\large}{15}{15}{}{}
\begin{center}
  \textbf{\LARGE Supplementary Information: Enhanced Sampling of Configuration and Path Space in a Generalized Ensemble by Shooting Point Exchange}\\[.2cm]
  Sebastian Falkner,$^{1}$ Alessandro Coretti,$^{1}$ and Christoph Dellago$^{1,*}$\\[.1cm]
  {\itshape ${}^1$University of Vienna, Faculty of Physics, 1090 Vienna, Austria.\\
  }
  ${}^*$Electronic address: christoph.dellago@univie.ac.at\\
(Dated: \today)\\[2cm]
\end{center}

\setcounter{equation}{0}
\setcounter{figure}{0}
\setcounter{table}{0}
\setcounter{page}{1}
\setcounter{section}{0}
\renewcommand{\theequation}{S\arabic{equation}}
\renewcommand{\thefigure}{S\arabic{figure}}
\renewcommand{\thetable}{S\arabic{table}}
\renewcommand{\bibnumfmt}[1]{[S#1]}
\renewcommand{\citenumfont}[1]{S#1}
\renewcommand{\thesection}{S\Roman{section}}
\renewcommand{\thepage}{S\arabic{page}}

\titleformat*{\section}{\Large\bfseries}

\section{Simulation Details for the Double Well System}

The double well has the potential energy form:
\begin{align}
    U(x) = \alpha \biggl\{0.25 \bigl[(x^{(0)})^2 + (x^{(1)})^2 - 4)^2 + (x^{(1)})^2\bigl]\biggr\}
\end{align}
where $\alpha$ was adjusted to match the desired barrier height. All simulations were run using an underdamped Langevin integrator~\cite{SGoga2012} with a friction of $20$, a timestep of $0.01$, a mass of $1$ and $k_{\text{B}}T = 1$. In simulations with and without shooting point exchange (SPEx), well-tempered metadynamics was configured identically placing a Gaussian of width $0.25$ every $100$ steps. The initial Gaussian height and bias factor were adjusted according to $0.2 / 15 \times A_{\text{barrier}}$ and $1 + (10 / 15 \times A_{\text{barrier}})$ respectively, where $A_{\text{barrier}}$ is the barrier height. The current bias was stored on a grid with a bin width of $0.1$ times the Gaussian width. Simulations with shooting point exchange followed the protocol of attempting an exchange every one TPS trial and $100$ metadynamics steps.

\section{Simulation Details for the KPTP-Peptide}

We prepared a simulation box with an edge length of \SI{4.95}{\nano \meter} including the KPTP tetrapeptide solvated in TIP3P water~\cite{SJorgensen1983}. We added K$^+$ and Cl$^-$ ions up to a concentration of \SI{150}{\milli\mole\per\liter} to neutralize the box. All simulations were performed employing OpenMM~\cite{SEastman2017} and PLUMED~\cite{SBonomi2019}. We used a velocity Verlet with velocity randomization integrator~\cite{SSivak2014} for simulation in the NVT ensemble at \SI{310}{\kelvin}. The timestep was set to \SI{2}{\femto \second} and the friction to \SI{1}{\per \pico \second}. All hydrogen bonds were constrained and the center of mass motion was removed at each timestep. Electrostatic interactions were treated using PME and the non-bonded cutoff was set to \SI{1.2}{\nano \meter}.

For SPEx, a well-tempered metadynamics and transition path sampling simulation were run independently and exchanges between them were performed every one TPS trial and $2000$ metadynamics simulation steps. For metadynamics, we bias along the $\omega^\text{P2}$ and $\xi^\text{P2}$ as described in the main text with a bias factor of $20$. The Gaussian width in both dimensions was \SI{0.3}{\radian} and the initial height was set to \SI{2}{\kilo \joule \per \mole}. A Gaussian kernel was placed every $100$ integration steps.

For the path sampling simulations, we define stable states as a function of $\omega^\text{P2}$ and $\xi^\text{P2}$:
\begin{align}
    h_{\text{trans}}(x) &= 1\ \text{if}\ (\omega^\text{P2} - \pi)^2 + (\xi^\text{P2} - \pi)^2 < 0.0625\\
    h_{\text{cis}}(x) &= 1\ \text{if}\ (\omega^\text{P2})^2 + (\xi^\text{P2} - \pi)^2 < 0.0625
\end{align}
and set a maximum path length of \SI{2}{\nano \second}. Shooting points were selected based on the current bias potential as described in the main text and velocities were redrawn from a Maxwell-Boltzmann distribution after selection to decorrelate paths faster.

\section{Training Protocol for Committor Learning}

Each exchange trial provides data on the committor in form of a configuration $x$ and whether the simulation reached \textit{cis} or \textit{trans} (0 / 1) from this point. We use a neural network (see table~\ref{tbl:NN_architecture}) to predict the expected outcome based on this data. Instead of $x$, we provide a set of collective variables (see table~\ref{tbl:cv_attribution}) and the network output is restricted to the range [0, 1]. The collective variables are normalized by subtracting their mean and dividing by their respective standard deviations. Each torsion angle is then mapped on two scalars in form of the sine and cosine of its value to enforce periodicity of the network output. The training is performed in Pytorch using a loss function of the form:
\begin{align}
    L = \frac{1}{N} \sum_{i=0}^{N-1} \log [ \epsilon + e^{-\hat y_i  y_i}]
\end{align}
where $N$ is the number of samples in the batch, $\hat y_i$ is the predicted label and $y_i$ the reference label. With $\epsilon = 1$, the loss function would be equivalent to a soft-margin loss, however, we use $\epsilon = 2.5$ to reduce the penalty of a misclassification. These are common since the training set only includes zeros or ones but we aim to predict probabilities that lie in between both values. We train for $15$ epochs on $2.5\times10^5$ data points using the Adam optimizer with a learning rate of $0.005$, a weight decay of $0.0001$ and an exponential learning rate decay with a decay rate of $0.8$. For improved accuracy, we train an ensemble of $10$ models and average their committor prediction for the final result. Attributions (table~\ref{tbl:cv_attribution}) were assigned by feature permutation as described in Jung et al.~\cite{SJung2021}.

The symbolic regression was performed using gplearn on $2500$ data points comprised of the three most important features from the attribution analysis and their network-predicted committor value. Parameters for reproducing the genetic search are given in table~\ref{tbl:sym_reg_parameters}. The fitness function applies a sigmoid to all search results and calculates the mean squared error of the prediction with respect to the target from the neural network output. We start $100$ independent optimization runs which we find more efficient in exploration than a larger population size or smaller tournament size. The ten best-performing models from these symbolic regression runs are shown in table~\ref{tbl:sym_reg_results}.


\begin{table}[htbp]
  \caption{Neural network architecture for the prediction of the committor in the KPTP tetrapeptide system.}
  \label{tbl:NN_architecture}
  \begin{tabular}{ll}
    \hline
    Layer & Components\\
    \hline
    Input Layer                               & Batch Norm  \\
                                              & Linear N$_\text{CVs}$ $\to$ 128 \\
                                              & Batch Norm + ReLU Activation \\
    &\\
    Layer 1                                   & Linear 128 $\to$ 64 \\
                                              & Batch Norm + ReLU Activation \\
    &\\
    Layer 2                                   & Linear 64 $\to$ 32 \\
                                              & Batch Norm + ReLU Activation \\
    &\\
    Layer 3                                   & Linear 32 $\to$ 16 \\
                                              & Batch Norm + ReLU Activation \\
    &\\
    Layer 4                                   & Linear 16 $\to$ 8 \\
                                              & Batch Norm + ReLU Activation \\
    &\\
    Output Layer                              & Linear 8 $\to$ 1 \\
                                              & Sigmoid \\
    \hline
  \end{tabular}
\end{table}

\begin{table}[htbp]
  \caption{Collective variable descriptors and their attribution score for the neural network-based prediction of the committor. On the right, the features are grouped by type (e.g. distance, angle, ...) and on the left they are sorted by their absolute attribution score. If no reference is given, notations are as follows: $d$ distance, $\theta$ angle, $\varphi$ torsion angle, [$\Phi, \Psi, \omega, \chi$] backbone/sidechain torsion angles.}
  \label{tbl:cv_attribution}
  \footnotesize
  \begin{tabular}{lr|lr}
    \hline
    Collective Variable & Attribution & Collective Variable & Attribution\\
    \hline
$d^{\text{K1N}_\zeta, \text{T3O}_\gamma}$                                   & 2.85E-03 & $\zeta^\text{P2}$ \cite{Fischer1994}                                                   & 1.80E-01\\
$d^{\text{K1O}, \text{K1N}}$                                                & -1.10E-04 & $\Omega^\text{P2}$                                                          & 1.05E-01\\
$d^{\text{K1N}_\zeta,\text{C}_\text{ter}}$                                  & 3.38E-03 & $\omega^\text{P2}$                                                          & 7.91E-02\\
$d^{\text{K1O}, \text{ACE-C}}$                                              & -2.29E-05 & $R_g$                                                                       & 4.05E-02\\
$d^{\text{K1O}, \text{ACE-O}}$                                              & -6.92E-04 & $\Phi^\text{P2}$                                                            & 1.43E-02\\
$d^{\text{K1O}, \text{K1N}}$                                                & 2.10E-04 & $\Psi^\text{P2}$                                                            & 8.76E-03\\
$d^{\text{K1O}, \text{K1C}_\alpha}$                                         & -3.44E-05 & $d^{\text{K1O}, \text{T3C}_\alpha}$                                         & 7.26E-03\\
$d^{\text{K1O}, \text{K1C}_\beta}$                                          & -3.66E-04 & $d^{\text{end-end}}$                                                        & 3.46E-03\\
$d^{\text{K1O}, \text{K1C}_\gamma}$                                         & 3.24E-08 & $d^{\text{K1N}_\zeta,\text{C}_\text{ter}}$                                  & 3.38E-03\\
$d^{\text{K1O}, \text{K1C}_\delta}$                                         & 5.13E-04 & $\chi_1^\text{K1}$                                                          & 3.31E-03\\
$d^{\text{K1O}, \text{K1C}_\epsilon}$                                       & -5.08E-08 & $d^{\text{K1N}_\zeta, \text{T3O}_\gamma}$                                   & 2.85E-03\\
$d^{\text{K1O}, \text{K1C}}$                                                & -6.35E-05 & $d^{\text{K1O}, \text{P2O}}$                                                & 1.97E-03\\
$d^{\text{K1O}, \text{P2N}}$                                                & -4.64E-05 & $\varphi^{\text{K1O, K1C, K1C}_\alpha \text{, K1C}_\beta}$                  & -1.61E-03\\
$d^{\text{K1O}, \text{P2C}_\delta}$                                         & -2.74E-06 & Puckering Phase \cite{Huang2014}                                                     & -1.40E-03\\
$d^{\text{K1O}, \text{P2C}_\gamma}$                                         & 2.46E-04 & $\chi_4^\text{P2}$                                                          & -1.23E-03\\
$d^{\text{K1O}, \text{P2C}_\beta}$                                          & -2.56E-04 & $\chi_1^\text{P2}$                                                          & 7.53E-04\\
$d^{\text{K1O}, \text{P2C}_\alpha}$                                         & 8.64E-07 & $\eta^\text{P2}$ \cite{Fischer1994}                                                    & 7.49E-04\\
$d^{\text{K1O}, \text{P2C}}$                                                & 1.27E-07 & $\chi_4^\text{K1}$                                                          & 7.24E-04\\
$d^{\text{K1O}, \text{P2O}}$                                                & 1.97E-03 & $d^{\text{K1O}, \text{ACE-O}}$                                              & -6.92E-04\\
$d^{\text{K1O}, \text{T3N}}$                                                & -2.79E-05 & $\Phi^\text{T3}$                                                            & 6.39E-04\\
$d^{\text{K1O}, \text{T3C}_\alpha}$                                         & 7.26E-03 & $d^{\text{K1O}, \text{K1C}_\delta}$                                         & 5.13E-04\\
$\theta^{\text{K1O},\text{K1C}, \text{K1C}_\alpha}$                         & -3.98E-06 & Puckering Amplitude \cite{Huang2014}                                                 & -5.07E-04\\
$\theta^{\text{K1C}_\alpha, \text{K1C}, \text{P2N} }$                       & 1.81E-04 & $\theta^{\text{K1C}, \text{P2N}, \text{P2C}_\alpha, }$                      & -4.84E-04\\
$\theta^{\text{K1C}, \text{P2N}, \text{P2C}_\alpha, }$                      & -4.84E-04 & $\omega^\text{T3}$                                                          & -4.07E-04\\
$\varphi^{\text{K1O, K1C, K1C}_\alpha \text{, K1N}}$                        & -1.06E-04 & $d^{\text{K1O}, \text{K1C}_\beta}$                                          & -3.66E-04\\
$\varphi^{\text{K1O, K1C, K1C}_\alpha \text{, K1C}_\beta}$                  & -1.61E-03 & $\omega^\text{P4}$                                                          & 3.47E-04\\
$\Omega^\text{P2}$                                                          & 1.05E-01 & Puckering $Z_y$ \cite{Huang2014}                                                     & -2.60E-04\\
$\varphi^\text{K1O, K1C, P2N, P2C}_\alpha$                                  & 9.16E-08 & $d^{\text{K1O}, \text{P2C}_\beta}$                                          & -2.56E-04\\
$\varphi^\text{ACE}$                                                        & -5.50E-05 & $d^{\text{K1O}, \text{P2C}_\gamma}$                                         & 2.46E-04\\
$\omega^\text{K1}$                                                          & -3.63E-05 & $\Psi^\text{T3}$                                                            & 2.40E-04\\
$\Psi^\text{K1}$                                                            & 9.98E-05 & $\chi_2^\text{P2}$                                                          & 2.27E-04\\
$\Phi^\text{K1}$                                                            & 1.32E-04 & $d^{\text{K1O}, \text{K1N}}$                                                & 2.10E-04\\
$\chi_1^\text{K1}$                                                          & 3.31E-03 & $\theta^{\text{K1C}_\alpha, \text{K1C}, \text{P2N} }$                       & 1.81E-04\\
$\chi_2^\text{K1}$                                                          & -8.33E-05 & $\xi^\text{P2}$                                                             & -1.61E-04\\
$\chi_3^\text{K1}$                                                          & -1.48E-04 & $\chi_4^\text{P4}$                                                          & -1.51E-04\\
$\chi_4^\text{K1}$                                                          & 7.24E-04 & $\chi_3^\text{K1}$                                                          & -1.48E-04\\
$\omega^\text{P2}$                                                          & 7.91E-02 & $\Phi^\text{K1}$                                                            & 1.32E-04\\
$\Psi^\text{P2}$                                                            & 8.76E-03 & $\chi_1^\text{T3}$                                                          & 1.26E-04\\
$\Phi^\text{P2}$                                                            & 1.43E-02 & $d^{\text{K1O}, \text{K1N}}$                                                & -1.10E-04\\
$\chi_1^\text{P2}$                                                          & 7.53E-04 & $\varphi^{\text{K1O, K1C, K1C}_\alpha \text{, K1N}}$                        & -1.06E-04\\
$\chi_2^\text{P2}$                                                          & 2.27E-04 & $\Psi^\text{K1}$                                                            & 9.98E-05\\
$\chi_3^\text{P2}$                                                          & 6.49E-06 & $\varphi^\text{NME}$                                                        & -9.48E-05\\
$\chi_4^\text{P2}$                                                          & -1.23E-03 & $\chi_2^\text{K1}$                                                          & -8.33E-05\\
$\xi^\text{P2}$                                                             & -1.61E-04 & $\Phi^\text{P4}$                                                            & -7.67E-05\\
$\omega^\text{T3}$                                                          & -4.07E-04 & $\Psi^\text{P4}$                                                            & -6.91E-05\\
$\Psi^\text{T3}$                                                            & 2.40E-04 & $d^{\text{K1O}, \text{K1C}}$                                                & -6.35E-05\\
$\Phi^\text{T3}$                                                            & 6.39E-04 & $\chi_2^\text{P4}$                                                          & -5.63E-05\\
$\chi_1^\text{T3}$                                                          & 1.26E-04 & $\varphi^\text{ACE}$                                                        & -5.50E-05\\
$\omega^\text{P4}$                                                          & 3.47E-04 & $d^{\text{K1O}, \text{P2N}}$                                                & -4.64E-05\\
$\Psi^\text{P4}$                                                            & -6.91E-05 & $\chi_1^\text{P4}$                                                          & -3.93E-05\\
$\Phi^\text{P4}$                                                            & -7.67E-05 & $\omega^\text{K1}$                                                          & -3.63E-05\\
$\chi_1^\text{P4}$                                                          & -3.93E-05 & $d^{\text{K1O}, \text{K1C}_\alpha}$                                         & -3.44E-05\\
$\chi_2^\text{P4}$                                                          & -5.63E-05 & $d^{\text{K1O}, \text{T3N}}$                                                & -2.79E-05\\
$\chi_3^\text{P4}$                                                          & -9.88E-06 & $d^{\text{K1O}, \text{ACE-C}}$                                              & -2.29E-05\\
$\chi_4^\text{P4}$                                                          & -1.51E-04 & $\chi_3^\text{P4}$                                                          & -9.88E-06\\
$\varphi^\text{NME}$                                                        & -9.48E-05 & $\chi_3^\text{P2}$                                                          & 6.49E-06\\
$\zeta^\text{P2}$ \cite{Fischer1994}                                                   & 1.80E-01 & $\theta^{\text{K1O},\text{K1C}, \text{K1C}_\alpha}$                         & -3.98E-06\\
$\eta^\text{P2}$ \cite{Fischer1994}                                                    & 7.49E-04 & $d^{\text{K1O}, \text{P2C}_\delta}$                                         & -2.74E-06\\
Puckering Phase \cite{Huang2014}                                                     & -1.40E-03 & $d^{\text{K1O}, \text{P2C}_\alpha}$                                         & 8.64E-07\\
Puckering Amplitude \cite{Huang2014}                                                 & -5.07E-04 & Puckering $Z_x$ \cite{Huang2014}                                                     & 2.54E-07\\
Puckering $Z_x$ \cite{Huang2014}                                                     & 2.54E-07 & $d^{\text{K1O}, \text{P2C}}$                                                & 1.27E-07\\
Puckering $Z_y$ \cite{Huang2014}                                                     & -2.60E-04 & $\varphi^\text{K1O, K1C, P2N, P2C}_\alpha$                                  & 9.16E-08\\
N$_\text{H}$ 0.5 nm around K1O                                              & -5.13E-08 & N$_\text{H}$ 0.5 nm around K1O                                              & -5.13E-08\\
$R_g$                                                                       & 4.05E-02 & $d^{\text{K1O}, \text{K1C}_\epsilon}$                                       & -5.08E-08\\
$d^{\text{end-end}}$                                                        & 3.46E-03 & $d^{\text{K1O}, \text{K1C}_\gamma}$                                         & 3.24E-08\\
    \hline
  \end{tabular}
\end{table}

\begin{table}[htbp]
  \caption{Parameters for the symbolic regression of the committor function.}
  \label{tbl:sym_reg_parameters}
  \begin{tabular}{ll}
    \hline
    Parameter & Value\\
    \hline
    Population Size & $1000$ \\
    Tournament Size & $20$ \\
    Constant Range & $-2\pi$ - $2\pi$ \\
    Generations & $100$ \\
    $P_\text{crossover}$ & $0.7$ \\
    $P_\text{subtree-mutation}$ & $0.1$ \\
    $P_\text{hoist-mutation}$ & $0.05$ \\
    $P_\text{point-mutation}$ & $0.1$\\
    Parsimony Coefficient  & $0.0001$ \\
    Function Set & Add, Sub., Div., Mul., Sin, Cos \\
    \hline
  \end{tabular}
\end{table}

\begin{table}[htbp]
  \caption{Symbolic regression results for the committor prediction with the lowest error.}
  \label{tbl:sym_reg_results}
  \begin{tabular}{lr}
    \hline
    Symbolic Regression Result & Mean Square Error\\
    \hline
$- \sin{\left(\Omega^\text{P2} - 0.75 \right)} + 4.334 \cos{\left(\zeta^\text{P2} \right)} + \cos{\left(\Omega^\text{P2} \right)} - 0.635                                                        $ & 3.201E-02 \\
$- 1.62 (\zeta^\text{P2})^{2} + \cos{\left(\Omega^\text{P2} \right)} + \cos{\left(\Omega^\text{P2} - 5.266 \right)} + 3.354                                                                        $ & 3.205E-02 \\
$\sin{\left(\Omega^\text{P2} + 2.474 \right)} + 4.703 \cos{\left(\zeta^\text{P2} \right)} + \cos{\left(\Omega^\text{P2} \right)} - 0.743                                                         $ & 3.240E-02 \\
$\sin{\left(\Omega^\text{P2} + 2.497 \right)} + 4.778 \cos{\left(\zeta^\text{P2} \right)} + \cos{\left(\cos{\left(\Omega^\text{P2} \right)} + 4.099 \right)}                                     $ & 3.249E-02 \\
$3.5308 \cos{\left(\Omega^\text{P2} \right)} + 2 \cos{\left(\omega^\text{P2} \right)} + \cos{\left(\zeta^\text{P2} - 5.354 \right)} - 0.283                                                      $ & 3.255E-02 \\
$4.706 \sin{\left(\zeta^\text{P2} - 4.62 \right)} + \cos{\left(\Omega^\text{P2} \right)} + \cos{\left(\Omega^\text{P2} + 0.466 \right)} - 0.682                                                  $ & 3.260E-02 \\
$\sin{\left(\Omega^\text{P2} + 2.387 \right)} + 4.658 \cos{\left(\zeta^\text{P2} \right)} + \cos{\left(\Omega^\text{P2} \right)} - 0.74                                                          $ & 3.277E-02 \\
$- (\zeta^\text{P2})^{2} - \sin{\left(\Omega^\text{P2} \right)} + \cos{\left(\zeta^\text{P2} \right)} + 2 \cos{\left(\Omega^\text{P2} \right)} + 1.918                                                       $ & 3.281E-02 \\
$- 0.889 \sin{\left(\Omega^\text{P2} \right)} + 3.458 \cos{\left(\Omega^\text{P2} \right)} + 2.186 \cos{\left(\omega^\text{P2} \right)}                                                           $ & 3.304E-02 \\
$- (\Omega^\text{P2})^{2} + \sin{\left(\zeta^\text{P2} - 3.792 \right)} + \cos{\left(\zeta^\text{P2} \right)} + \cos{\left(\omega^\text{P2} \right)} + \cos{\left(\cos{\left(\omega^\text{P2} \right)} \right)} + 1.366 $ & 3.319E-02 \\

    \hline
  \end{tabular}
\end{table}

\end{document}